\documentclass[twocolumn,prl,showpacs,showkeywords]{revtex4}
\usepackage{graphicx}
\begin{document}

\title{Dyson Orbitals, Quasi-Particle effects and Compton scattering}

\author{B. Barbiellini and A. Bansil}

\address{Department of Physics, Northeastern University, Boston, 
         MA 02115 USA}


\pacs{78.70.Ck, 71.10Ca, 31.25.Eb}
\keywords{Green's function,
Density Matrix, 
Electron-Electron Correlation, 
Dyson Orbitals, Electron Momentum Density, 
Compton Profile}
\begin{abstract}

Dyson orbitals play an important role in understanding quasi-particle
effects in the correlated ground state of a many-particle system and are
relevant for describing the Compton scattering cross section beyond the
frameworks of the impulse approximation (IA) and the independent particle
model (IPM). Here we discuss corrections to the Kohn-Sham energies due to
quasi-particle effects in terms of Dyson orbitals and obtain a relatively
simple local form of the exchange-correlation energy. Illustrative
examples are presented to show the usefulness of our scheme.

\end{abstract}
\maketitle

\section{I. Introduction}

Dyson orbitals are a set of one-particle orbitals that are associated with
many-electron wavefunctions. These orbitals connect the exact ground-state
of the $N$-electron system with excited states containing $N-1$ or $N+1$
electrons. The importance of Dyson orbitals in understanding Compton
scattering spectra has been emphasized recently by Kaplan et {\em al.}
\cite{kaplan}, who present a general formalism for the Compton cross
section, which goes beyond the standard treatment involving the frameworks
of the impulse
approximation (IA) and the independent particle model (IPM). 
The breakdown of the IA in describing core Compton
profiles is well-known \cite{eis70}. More recently, high
resolution valence Compton profiles (CPs) of Li at relatively low photon
energy of $8-9$ keV have been found to show asymmetries in shape
and smearing of the Fermi surface (FS) features where
deviations from the IA have been implicated \cite{sternemann,soininen}.

Dyson orbitals also give insight into quasi-particle effects in the
correlated ground state of the many body system. In this article, we focus
on understanding energies of one-particle excitations of the ground
state, which play an important role not only in the formalism of the
Compton scattering cross-section, but also in the band structure problem
more generally. To this end, a Green's function approach is used to first
obtain an expression for the exchange-correlation energy in terms of the
self-energy operator involved in the description of the Dyson orbitals. A
local ansatz for the self-energy is then invoked to obtain a relatively
simple expression for the excitation energies. We illustrate our scheme by
considering the example of first ionization energies of low $Z$ atoms from
$Z=1$ (H) to $Z=6$ (C) and find good agreement with the corresponding
experimental results. As another example, the measured bandgap in diamond
is also reproduced reasonably by our computations. 

An outline of this article is as follows. The introductory remarks are
followed in Section II by a brief overview of the general formalism of
Ref. \cite{kaplan} for the Compton scattering cross section. The importance of
properly including excitation energies in the computation for describing
the asymmetry of the CP around $q=0$ is stressed. Section III presents the
Green's function formulation and discusses our scheme for computing
excitation energies. Section IV gives a few illustrative applications of
the theory, followed in Section V by a few concluding remarks.

\section{II. Dyson orbitals and Compton scattering}

The Dyson spin-orbital $g_n$ can  
be defined in terms of the many-body ground-state
wavefunction $\Psi_{0}$ and the wavefunction 
$\Psi_{n}$ of the singly ionized excited system
characterized by the quantum number $n$ \cite{43,43bis,44,45}:
\begin{equation}
g_{n}(\mathbf{x}_{N})=
\sqrt{N}\int~\Psi_{n}(\mathbf{x}_{1}....
\mathbf{x}_{N-1})^{*}\Psi_{0}(\mathbf{x}_{1}
....\mathbf{x}_{N})~d\mathbf{x}_{1}....
d\mathbf{x}_{N-1}~, 
\label{eq12} 
\end{equation}
where the index $\bf x$ denotes both spatial and spin coordinates and 
the integration over $d\mathbf{x}_{i}$ implicitly includes a 
summation over the spin coordinates. The Dyson 
orbitals thus give 
generalized overlap amplitudes between the ground state 
and the singly ionized states of the many body system. 
Note that, in general, Dyson orbitals do
not form an orthonormal set. The Dyson spin-orbital with
the spin projection $\sigma$ may be written in terms of the 
spin function $\sigma(\zeta)$ as
\begin{equation}
g_{n}(\mathbf{x})=g_{n}(\mathbf{r},\sigma({\zeta}))
=g_{n}(\mathbf{r})\sigma(\zeta)~.  
\label{eq13}
\end{equation}
The excitation energy $E_{b}^{(n)}$ associated with the $n^{th}$ Dyson
orbital is given by
\begin{equation}
E_{b}^{(n)}=E_{n}(N-1)-E_{0}(N)  
\label{eq2}
\end{equation}
where $E_{0}(N)$ is the $N$-particle ground state energy 
and $E_{n}(N-1)$ is the energy of the $(N-1)$ particle 
ionized system when it is in its $n^{th}$ quantum state. 
Eqs. 1-3 above define what may be thought of as 
occupied Dyson orbitals. For completeness, one can also 
introduce a parallel set of
"unoccupied" Dyson orbitals, which connect the ground state 
to various $(N+1)$ particle states containing an added electron.  

The importance of Dyson orbitals in understanding the nature of the
Compton scattering spectra beyond the framework of the 
IA and the IPM has been
emphasized by Kaplan {\em et al.} \cite{kaplan}. In particular, the
triple-differential scattering cross-section 
for the ($\gamma$,$e\gamma$) process, 
which is the elementary process underlying Compton
scattering, is given by
\begin{eqnarray} \frac{d^{3}\sigma}{d\omega_{2}d\Omega_{2}d\Omega_{e}}=
r_{0}^{2}(1+\cos^{2}\theta)\frac{\omega_{2}}{\omega_{1}}\nonumber \\
\times\sum_{n}|g_{n}({\mathbf{q}})|^{2}
\delta(\omega_{1}-\omega_{2}-E_{b}^{(n)}-\frac{p_{n}^{2}}{2m})~,
\label{eqcross}
\end{eqnarray} 
where $g_{n}({\mathbf{q}})$ is the Fourier transform of $g_n({\bf r})$,
$\mathbf{q}$ is the momentum transfered to the final system, $\omega_{1}$
and $\omega_{2}$ are respectively the energies of the photon before and
after the collision, and the summation extends over the occupied Dyson
orbitals. Eq. (\ref{eqcross}) assumes a large energy transfer
($\omega_{1}-\omega_{2}>>E_{b}^{(n)}$), so that the outgoing electron
possesses sufficiently high energy that it can be approximated by a free
electron plane wave form. The binding energy $E_{b}^{(n)}$ in the
$\delta$-function in Eq. (\ref{eqcross}) is usually neglected in
obtaining the standard IA-based expression for the cross section. The
resulting form of the Compton profile after coordinates of the recoil
electron are integrated over can be shown to be symmetric around $q=0$.
Therefore, an accurate computation of the binding energies $E_{b}^{(n)}$
is important for describing the asymmetry of the Compton profile related
to deviations from the IA.

\section{III. Green's function method}

In order to gain a handle on the nature of the excitation energies
$E_{b}^{(n)}$, it proves useful to approach the problem through the
Green's function method. The orbitals $g_n({\bf r})$ 
of Eq. (\ref{eq12}) satisfy the
Dyson equation \cite{dyson1,dyson2,mazin}
\begin{eqnarray}
(\frac{p^2}{2m}+V_{ext}({\bf r})+V_H({\bf r}))g_n({\bf r})
\nonumber \\
+ \int d^3{\bf r'} ~\Sigma_{xc}({\bf r},{\bf r'},E_{b}^{(n)})
g_n({\bf r'})
=E_{b}^{(n)}g_n({\bf r}),
\end{eqnarray}
where $V_{ext}({\bf r})$ is the external potential, $V_H$ 
is the Hartree
potential and $\Sigma_{xc}$ is the self-energy. 

The Green's function can be expressed as \cite{mazin}
\begin{equation}
G({\bf r},{\bf r'},\omega)=\sum_n
\frac{g_{n}({\bf r})g_{n}^*(\bf r')}{\omega-E_{b}^{(n)}+
i\delta \mbox{sign}(E_{b}^{(n)}-\mu)},
\label{eq_green}
\end{equation}
where $\mu$ is the chemical potential and $\delta$ is 
an infinitesimal positive number.
The density matrix $\rho({\bf r},{\bf r'})$ and the electron density
distribution $n({\bf r})=\rho({\bf r},{\bf r})$ are obtained by
integrating the spectral function 
$A({\bf r},{\bf r'},\omega)$
over the occupied electronic energies 
\begin{equation} 
\rho({\bf r},{\bf r'})=\int_{-\infty}^{\mu} 
d\omega A({\bf r},{\bf r'},\omega)~,
\end{equation} 
where 
the spectral function  
$A({\bf r},{\bf r'},\omega)$ is given by
\begin{equation} 
A({\bf r},{\bf r'},\omega)=\sum_n g_{n}({\bf r})g_{n}^*({\bf r'}) 
\frac{\delta}{\pi[(\omega-E_{b}^{(n)})^2+\delta^2]}.
\end{equation} 

The standard calculation of the self-energy $\Sigma_{xc}$ and the Green's
function $G$ proceeds via the many-body perturbation theory (MBPT)
\cite{kheifets01}. The first order in the MBPT leads to the so-called $GW$
approximation \cite{hedin65} ($G$ stands for the Green's function and $W$
denotes the screened Coulomb interaction). The $GW$ equation for the
self-energy is 
\begin{equation} \Sigma_{xc}= iG_0 W~, \label{GW}
\end{equation} 
where $G_0$ is the 0$^{th}$ order Green's function. The calculation of
$W$
usually requires heavy computations of the dielectric function. Here, we
will take a shortcut by describing the screening in terms of the
pair-correlation function.

Since many-body effects beyond the Hartree approximation are contained in
the exchange-correlation energy $E_{xc}$, we can write
\begin{equation}
E_{xc}=\frac{1}{2}\int_{-\infty}^{\mu} d\omega d^3{\bf r} d^3{\bf r'}
\Sigma_{xc}({\bf r},{\bf r'},\omega)A({\bf r},{\bf r'},\omega)~,
\label{eq_xca}
\end{equation}
where the integral is over the occupied states. In the Hartree-Fock limit,
the self energy operator is {\em instantenous}, i.e., 
$\Sigma_{xc}({\bf r},{\bf r'},\omega)= \Sigma_{x}({\bf r},{\bf r'})$. 
By integrating over $\omega$, ones finds the well known result for the
exchange energy $E_{x}$ \cite{ashcroft}
\begin{equation}
E_{x}=\frac{1}{2}\int d^3{\bf r} d^3{\bf r'}
\Sigma_{x}({\bf r},{\bf r'})\rho({\bf r},{\bf r'})~.
\end{equation}

We now consider the standard expression of the exchange-correlation energy
$E_{xc}$ in terms of the exchange-correlation hole \cite{reviewdft}
\begin{equation}
E_{xc}=\frac{1}{2}
\int d^3{\bf r} d^3{\bf r'} ~ 
\int_0^1 d\lambda \frac{n({\bf r})n({\bf r'})
C_{\lambda}({\bf r},{\bf r'})}
{|{\bf r}-{\bf r'}|}~,
\label{eq_xchole}
\end{equation}
where $\lambda$ is the coupling constant from the Hellmann-Feynman
theorem, and $C_{\lambda}$ is a pair-correlation function describing the
exchange-correlation hole. 
The two expressions 
of Eqs. (\ref{eq_xca}) and (\ref{eq_xchole}) for
$E_{xc}$ can be linked by assuming 
a {\em local} and {\em instantaneous}
ansatz for the self-energy operator \cite{kulikov87}
\begin{equation}
\Sigma_{xc}({\bf r},{\bf r'},\omega)=
2\epsilon_{xc}({\bf r})\delta({\bf r}-{\bf r'}).
\label{eq_sigma}
\end{equation}
Here
\begin{equation}
\epsilon_{xc}({\bf r})=
\frac{1}{2}
\int d{\bf r'} \int_0^1 d\lambda \frac{n({\bf r})n({\bf r'})
C_{\lambda}({\bf r},{\bf r'})}
{|{\bf r}-{\bf r'}|}.
\end{equation}
is the exchange-correlation energy per particle. 

Eq.(\ref{eq_sigma}) has an intuitive interpretation. When an electron
moves, its exchange-correlation hole moves with it and modifies its
effective mass and excitation energy. Insight into the local approximation
for $\Sigma_{xc}$ is provided by considering the homogeneous electron gas.  
In this case, one obtains
\begin{equation}
\Sigma_{xc}({\bf r},{\bf r'},\omega)=
2{\bar \epsilon_{xc}}\delta({\bf r}-{\bf r'}),
\end{equation}
where ${\bar \epsilon_{xc}}$ is a constant 
due to the translation invariance of
the system.  
A simple expression 
for ${\bar \epsilon_{xc}}$ in the metallic density range
is given by \cite{bba89}
\begin{equation}
{\bar \epsilon_{xc}}=
-\frac{0.916}{r_s}
-\frac{0.127}{ \sqrt{r_{s}}}~\mbox{Ry},
\end{equation}
where the mean interelectronic spacing $r_{s}$
is obtained via the electron density
$n$ by $n (4\pi r_{s}^{3}/3)=1$.
Thus, as a consequence of the short range of the self-energy,
the electron-electron interaction leads only to a uniform shift of the
levels in relation to the non-interacting gas.  In the Hartree-Fock case,
the density of states unphysically goes to zero at the Fermi level as a
result of the long range of the self-energy \cite{ashcroft}. The situation
in a real metal is presumably closer to that of the homogeneous electron
gas as the screening tends to make the correct $\Sigma_{xc}$ local.
Indeed, band calculations based on the 
Density Functional Theory (DFT) employing local potentials
\cite{reviewdft} have been
rather successful in reproducing the 
experimental FS's in wide
classes of materials \cite{bandrev}.

Bearing these considerations in mind, 
the correction to the excitation
energy $E_{b}^{(n)}$ of Eq. (\ref{eq_green}) may be obtained in the first
perturbational order as
\begin{equation}
E_{b}^{(n)}=\varepsilon_n
+\int d^3 {\bf r}~ 
(2\epsilon_{xc}({\bf r})-v_{xc}({\bf r}))|g_n({\bf r})|^2~,
\label{eq_excite}
\end{equation}
where $\varepsilon_n$ denotes the Kohn-Sham eigenvalue and $v_{xc}({\bf
r})=\delta E_{xc}/\delta n({\bf r})$ is the exchange-correlation potential
in the Kohn-Sham equations. $g_n(\bf r)$ on the right hand side 
of Eq. (\ref{eq_excite})
can be reasonably replaced by the Kohn-Sham orbitals
\cite{duffy94}. 
The exchange-correlation energy per particle
$\epsilon_{xc}$ is computed straightforwardly within the local density
approximation (LDA) \cite{reviewdft}.
Notably, correction to $E_b^{(n)}$ 
in Eq. (\ref{eq_excite}) will in
general be state-dependent and therefore this spectrum will be a useful
starting point for implementing the scheme proposed by Barbiellini and
Bansil \cite{agp} for treating the momentum density and Compton profile of
the correlated ground state of the anisotropic electron gas beyond the
Lam-Platzman correction. 

\section{IV. Excitation energy calculations}

\begin{figure}
\begin{center}
\includegraphics[width=\hsize]{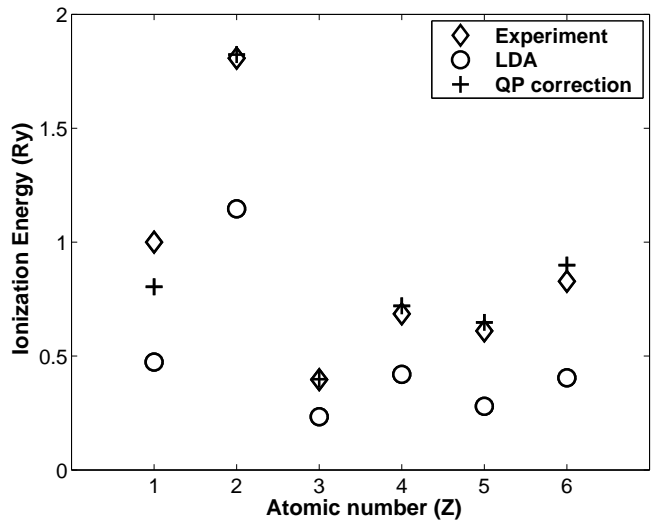}
\end{center}
\caption{
Comparison of the theoretical and experimental atomic first ionization
energies for H ($Z=1$) to C ($Z=6$). 
Diamonds are the experimental values \cite{pauling},
open circles give LDA values,
while crosses give values corrected by using 
Eq. (\ref{eq_excite}).
}
\label{fig1}
\end{figure}

Fig. \ref{fig1} provides an illustrative example of the usefulness 
of Eq. (\ref{eq_excite}).
The first ionization energy of atoms from H to C ($Z=1-6$) is
considered using relativistic DFT atomic wavefunctions \cite{relat}. The
exchange-correlation energy $\epsilon_{xc}({\bf r})$ and the potential
$v_{xc}({\bf r})$ have been calculated within the
LDA parametrized by Hedin and Lundqvist \cite{lda}. The
LDA eigenvalues (open circles) are seen to be substantially lower than the
experimental values (diamonds).  The quasiparticle correction of 
Eq. (\ref{eq_excite}) brings the
theoretical values (crosses) in substantial agreement with the
experimental values \cite{pauling}. 
Note that in the case of H, although the  
quasiparticle correction yields a large improvement, 
the self-interaction error is not fully removed
and about $20$ \% discrepancy between theory and
experiment remains. 
\

We have also applied our scheme to investigate the band gap in diamond.
For this purpose, the selfconsistent electronic band structure 
of diamond \cite{epmd-lmto}
was obtained using the local density exchange-correlation functional of
Hedin and Lundqvist \cite{lda}. As expected, the LDA band gap of $3.99$ eV
so computed is too small. The inclusion of the correction 
of Eq. (\ref{eq_excite})
yields a gap of $5.3$ eV, which is in far better agreement with the
experimental value of $5.48$ eV \cite{clark64}. 
More generally, we expect our
correction to reproduce semiconductor bandgaps with an accuracy comparable
to that of the computationally more involved $GW$ approximation
\cite{review_GW,hybertsen85,godby86,gygi89,zhu91}. It should also be noted
that in analyzing the generalized density-functional theory (GDFT),
several authors have pointed out that the energy of electronic excitations
across the gap of insulators and semiconductors can be expressed as the
sum of the so-called Kohn-Sham gap and a correction that is usually of the
same order of magnitude \cite{fritsche86,fritsche93,remed99}.

We note that quasi-particle effects directly influence the shape of the FS
of a metal. For instance, the FS of Cu can be measured very precisely 
so that direct comparisons with theoretical predictions are
easy to follow. de Hass-van Alphen (dHvA) measurements\cite{cudhva} find
the large belly of the FS of Cu to be more spherical compared to the LDA
computations. In V, dHvA \cite{vdhva} and positron annihilation
\cite{vpos} experiments indicate that the FS pockets at the N symmetry
point are smaller than those from LDA calculations. In $\gamma$-Ce, an
improved agreement with positron annihilation results is obtained with the
LDA FS if the $f$-band is moved up by $40-50$ mRy \cite{cepos}.  
Interestingly, aforementioned discrepancies concerning the FSs can be
substantially corrected via a simplified computation of the self-energy
via the exchange correlation hole \cite{bjhole}.

\section{V. Summary and Conclusions}
We discuss aspects of Dyson orbitals for gaining insight into
quasi-particle effects in the correlated ground state of a many-particle
system. The importance of Dyson orbitals, which connect the many-body
ground state with its singly ionized excited states, has been emphasized
previously for describing Compton scattering profiles beyond the
limitations of the IA and the IPM and we start with a brief review of this
earlier study \cite{kaplan}. We focus on delineating 
corrections to the Kohn-Sham
energies due to quasi-particle effects and show how one can gain a handle
on obtaining substantially improved excitation energies in molecules and
solids. For this purpose, a
Green's function approach is used to suggest 
a relatively simple local
form of the exchange-correlation energy. Illustrative applications of our
scheme indicate that our self-energy expression should provide a
reasonable description of excitation energies in metals and semiconductors
with an accuracy roughly comparable to that of the GW method.


This work is supported by the US Department of Energy contract
DE-AC03-76SF00098 and benefited from the allocation of supercomputer time
at NERSC and Northeastern University's Advanced Scientific Computation
Center (ASCC).


\begin{thebibliography}{99}
%
%
\bibitem{kaplan}
I. G. Kaplan, B. Barbiellini, A. Bansil, 
Phys. Rev. B {\bf 68}, 235104 (2003).
%
%
\bibitem{eis70}
See e.g. P. Eisenberger and P.M. Platzman,
Phys. Rev. A {\bf 2}, 415 (1970).
\bibitem{sternemann}C. Sternemann, 
K. H\"{a}m\"{a}l\"{a}inen, A. Kaprolat, A. Soininen,
G. D\"{o}ring, C.-C. Kao, S. Manninen, and W. Sch\"{u}lke, Phys.
Rev. B \textbf{62}, R7687 (2000). 
\bibitem{soininen}J. A. Soininen, K. H\"{a}m\"{a}l\"{a}inen, 
and S. Manninen, Phys.  Rev. B \textbf{64}, 125116 (2001). 
%
%
\bibitem{43} 
B.T. Pickup, Chem. Phys. 
\textbf{19}, 193 (1977).
\bibitem{43bis} 
L.S. Cederbaum and W. Domcke, 
Adv. Chem. Phys. \textbf{36}, 205 (1977).
\bibitem{44} 
Y. \"Ohrn and G. Born, Adv. Quant. Chem. 
\textbf{13}, 1 (1981).
\bibitem{45} 
J. V. Ortiz, 
\emph{in Computational Chemistry: Reviews of
Current Trends}, Vol. 2, 
J. Leszczynski. Ed., (World Scientific, Singapore,
1997), p.1
%
%
\bibitem{dyson1}
A.J. Layzer, Phys. Rev. {\bf 129}, 897 (1963). 
\bibitem{dyson2} 
L. Hedin and S. Lundqvist 
in {\em Solids State Physics}, ed. H. Ehrenreich,
D. Turnbull and F. Seitz, (Academic, New York) (1969), 
Vol. 23, p. 1.
\bibitem{mazin} I.I. Mazin, E.G. Maksimov,
S. Yu. Savrasov and Yu. A. Uspenskil,
Sov. Phys. Solid State {\bf 29}, 1516 (1987).
%
%
\bibitem{kheifets01}
A.S. Kheifets, M. Vos and E. Weigold,
Zeitschrift fur Physikalische Chemie 
{\bf 215}, 1323 (2001).
%
%
\bibitem{hedin65}
L. Hedin, Phys. Rev. {\bf 139}, A796 (1965).
%
%
\bibitem{ashcroft}
N.W. Ashcroft and N.D. Mermin, {\em Solid State Physics},
(Saunders College, Philadelphia, 1976).
%
%
\bibitem{reviewdft} 
R.O. Jones and O. Gunnarsson, 
Rev. Mod. Phys {\bf 61}, 689 (1989). 
%
%
\bibitem{kulikov87}
See also
N. I. Kulikov, M. Alouani, 
M. A. Khan and M. V. Magnitskaya, 
Phys. Rev. B {\bf 36}, 929 (1987).
%
%
\bibitem{bba89}
B. Barbiellini, Phys. Lett. A {\bf 134}, 330 (1989).  
%
%
\bibitem{bandrev} D.D. Koelling, 
Rep. Prog. Phys. {\bf 44}, 139 (1981).
%
%
\bibitem{duffy94}
P. Duffy, D. P. Chong, M. E. Casida, and D. R. Salahub,
Phys. Rev. A {\bf 50}, 4707 (1994)
argue that Dyson orbitals can be represented reasonably
well by the Kohn-Sham orbitals for momentum density 
calculations.
%
%
\bibitem{agp}
B. Barbiellini and A. Bansil, 
J. Phys. Chem. Sol. {\bf 62}, 2181 (2001).
%
%
\bibitem{relat} D. D. Koelling and B.N. Harmon,
J. Phys. C: Solid St. Phys. {\bf 10}, 3107 (1975);
Relativistic atomic code provided by M. Posternak.
\bibitem{lda}
L. Hedin and B.I. Lundqvist, J. Phys. C {\bf 4}, 2064 (1971).
%
%
\bibitem{pauling}
L. Pauling, {\em General Chemistry}, Dover (New York, 1970).
The ionization energies in eV are
13.60, 24.58, 5.39, 9.32, 8.30
and 11.26 for H, He, Li, Be, B and C respectively.
%
%
\bibitem{epmd-lmto}
B. Barbiellini, S.B. Dugdale and T. Jarlborg,
Computational Materials Science {\bf 28},
287 (2003).
%
%
\bibitem{clark64}
C.D. Clark, P. J. Dean, and P. V. Harris, 
Proc. Roy. Soc. (London) A{\bf 277}, 312 (1964).
%
%
\bibitem{review_GW}
F. Aryasetiawan and O. Gunnarsson, Rep. Prog. Phys. {\bf 61} , 237 (1998). 
\bibitem{hybertsen85}
M. S. Hybertsen and S. G. Louie, Phys. Rev. Lett. {\bf 55}, 1418 (1985); 
Phys. Rev. B {\bf 34}, 5390 (1986). 
\bibitem{godby86}
R. W. Godby, M. Schl\"uter and L. J. Sham, 
Phys. Rev. Lett. {\bf 56}, 2415 (1986); 
Phys. Rev. B {\bf 37}, 10 159 (1988). 
\bibitem{gygi89}
F. Gygi and A. Baldereschi, Phys. Rev. Lett. {\bf 62}, 
2160 (1989). 
\bibitem{zhu91}
X. Zhu and S. G. Louie, Phys. Rev. B {\bf 43}, 14 142 (1991).
%
%
\bibitem{fritsche86}
L. Fritsche, Phys. Rev. B {\bf 33}, 3976 (1986).
\bibitem{fritsche93}
L. Fritsche and Y. M. Guo,
Phys. Rev. B {\bf 48}, 4197 (1993).
\bibitem{remed99}
I. N. Remediakis and E. Kaxiras,
Phys. Rev. B {\bf 59}, 5536 (1999).
%
%
\bibitem{cudhva} P.T. Coleridge and I.M. Templeton, 
Phys. Rev. B {\bf 25}  7818 (1982).
%
%
\bibitem{vdhva} 
 R.D. Parker and M.H. Halloran,
 Phys. Rev. B {\bf 9}, 4130 (1974).
\bibitem{vpos} A. A. Manuel, 
Phys. Rev. Lett. {\bf 20}, 1525 (1982);
 A.K. Singh, A.A. Manuel, R.M. Singru, R. Sachot,
 E. Walker, P. Descouts and M. Peter, 
J. Phys. F.{\bf 15}, 237 (1985).
%
%
\bibitem{cepos} T. Jarlborg, A. A. Manuel, M. Peter,
D. Sanchez, A.K. Singh, J.-L. Stephan, E. Walker, W. Asmuss
and M. Hermann
in {\em Positron Annihilation}, Proceedings ICPA8, edited
by L. Dorikens {\em et al.} (World-Scientific, Singapore) (1989) p. 266.
%
%
\bibitem{bjhole}
B. Barbiellini and T. Jarlborg, 
J.Phys.: Condens. Matter {\bf 1}, 75 (1989).
\end{thebibliography}
\end{document}